\documentclass[twocolumn,prb]{revtex4}
\usepackage{graphicx}
\begin{document}


\title{Phenomenological Study of Decoherence in Solid-State Spin Qubits due to Nuclear Spin Diffusion}


\author{Michael J. Biercuk}
\email{michael.biercuk@sydney.edu.au}
\affiliation{School of Physics, The University of Sydney, NSW 2006 Australia}

\author{Hendrik Bluhm}
\email{hendrikb@physics.harvard.edu}
\affiliation{Department of Physics, Harvard University, Cambridge, MA 02138 USA}


\date{\today}

\begin{abstract}
We present a study of the prospects for coherence preservation in solid-state spin qubits using dynamical decoupling protocols.  Recent experiments have provided the first demonstrations of multipulse dynamical decoupling sequences in this qubit system, but quantitative analyses of potential coherence improvements have been hampered by a lack of concrete knowledge of the relevant noise processes.  We present simulations of qubit coherence under the application of arbitrary dynamical decoupling pulse sequences based on an experimentally validated semiclassical model.  This phenomenological approach bundles the details of underlying noise processes into a single experimentally relevant noise power spectral density.   Our results show that the dominant features of experimental measurements in a two-electron singlet-triplet spin qubit can be replicated using a $1/\omega^{2}$ noise power spectrum associated with nuclear-spin-flips in the host material.  Beginning with this validation we address the effects of nuclear programming, high-frequency nuclear-spin dynamics, and other high-frequency classical noise sources, with conjectures supported by physical arguments and microscopic calculations where relevant.  Our results provide expected performance bounds and identify diagnostic metrics that can be measured experimentally in order to better elucidate the underlying nuclear spin dynamics.

\end{abstract}


\maketitle

\section{Introduction}

\indent Spin qubits in semiconductor materials are a leading experimental realization of a controllable, scalable quantum system in the solid-state~\cite{Hanson_RMP}.  These devices hold numerous advantages relative to other qubit implementations; First, the use of a semiconductor platform provides benefits in terms of scaling -- leveraging decades of development in advanced micro- and nanofabrication, large-scale semiconductor-based systems appear technically feasible.  Second, experimental demonstrations of coherent control of semiconductor spin qubits allow operations in the nanosecond and sub-nanosecond range~\cite{Petta_Science2005, Vandersypen_Science_SO, Vandersypen_PRL_Rabi}, ultimately permitting rapid logic operations at the physical qubit level.  Finally, measured electron-spin relaxation times ($T_{1}$) in semiconductor nanostructures exceed several seconds~\cite{Zumbuhl_PRL_T1}, placing the limit on coherence time ($T_{2}^{(max)}=2T_{1}$) up to 10 orders of magnitude larger than demonstrated gate times, an important DiVincenzo criterion~\cite{DVCriteria_1998}.  
\\
\indent Many realizations of semiconductor spin qubits have been proposed and demonstrated experimentally~\cite{Hanson_RMP}.  Additionally, experimental control fidelity has progressed to the point where probing studies of the coherence limits of spins in semiconductors are possible~\cite{Biercuk_Nano2011}.  We focus on the two-electron Singlet-Triplet (S/T) spin qubit in GaAs~\cite{Petta_Science2005} as it represents one of the most advanced solid-state spin qubits available.  This system employs the symmetric and antisymmetric spin configurations of a pair of spin-$1/2$ electrons to form a qubit basis.  Electron pairs are confined in a lithographically pattered nanostructure based on a two-dimensional electron gas and confining electrostatic gates.  Decades of research in mesoscopic physics now permit experimentalists to isolate electron pairs, enact controlled exchange-based interactions, and perform a strong projective measurement via spin-to-charge conversion and use of a proximal charge detector~\cite{Reilly_APL_RFQPC}.  
\\
\indent The strengths of the S/T qubit, however, come at significant cost in terms of decoherence processes.  Most significantly, the localization of the qubit within a nanopatterned GaAs-heterostructure host material introduces noise sources due to fluctuating nuclear spins~\cite{Johnson_Nature2005, Taylor_PRB2007} and localized charge centers~\cite{Fujisawa_PRL_2003}.  A series of microscopic theoretical studies and experiments have shown that in these devices the effects of nuclear spin dynamics dominate measured qubit coherence.  To date, typical free-induction-decay (FID) $T_{2}^{(FID)}$ times have been measured $\sim$20-50 ns~\cite{Petta_Science2005, Reilly_Science_Nuke, Bluhm_Feedback}, and spin echo experiments (with exchange off)~\cite{Bluhm_LongCoherence} have yielded coherence times $T_{2}^{(Echo)}\approx30\;\mu$s.  Recent experiments have extended $T_{2}^{(FID)}$ to hundreds of ns using nuclear programming~\cite{Bluhm_Feedback}, and have demonstrated long coherence times using multipulse dynamical decoupling sequences~\cite{Barthel_Interlaced, Bluhm_LongCoherence} to mitigate the effects of dephasing.  At this time, however, an efficient and practical model providing quantitative insight into the ultimate prospects for the suppression of decoherence-induced error accumulation remains elusive. 
\\
\indent In this manuscript we use an experimentally validated theoretical model for error accumulation under the application of dynamical decoupling pulse sequences to accomplish four main goals: 1) Provide a general phenomenological framework for spin-qubit error accumulation in terms of noise power spectral densities by comparison with experimental measurements, 2) Provide insight into the spectral characteristics of dominant noise processes, 3)Understand the prospects for and limitations of dynamical decoupling in extending the coherence of spin qubits, and 4) Identify the most promising technical approaches to improving qubit coherence.  Our numerical simulations first suggest the coherence times achievable under dephasing due to fluctuating nuclear spins.  Calibrations of the noise power spectral density, $S_{\beta}(\omega)$, for angular frequency $\omega$, are performed using measurements of $T_{2}^{(FID)}$ and the relative coherence time for an $n$-pulse CPMG cycle relative to the FID time, $T_{2}^{(CPMG_{n})}/T_{2}^{(FID)}$.  Our calculations allow comparisons of the scaling of measured coherence times with pulse number, $n$, and studies of the influence of different spectral components of $S_{\beta}(\omega)$, particularly the form of the high-frequency cutoff.  Further, we are able to study predicted error rates under dynamical decoupling in the high-fidelity regime, vital for quantum computation. This model allows detailed comparison with experiment and provides predictive power for expected qubit coherence times in the presence of realistic noise sources.
\\
\indent The remainder of this manuscript is organized as follows:  Section~\ref{sec:Model} introduces the theoretical model used for the simulations and describes the relevant experiments.  This is followed by detailed simulations in Section~\ref{sec:1f2} of predicted coherence in the presence of a $1/\omega^{2}$ noise power spectral density.  More complex power spectra and the role of a high-frequency noise cutoff are introduced in Section~\ref{sec:White}.  The effects of noninstantaneous and imperfect control pulses are described in Section~\ref{sec:TauPi}, and the manuscript concludes with discussion in Section~\ref{sec:Discuss}.

\section{\label{sec:Model}Dephasing Noise in a Dynamical Decoupling Framework}
\indent We consider a simple theoretical model based on previous work  \cite{Uhrig2007,Uhrig2008,Lee2008, Cywinski2008}, and validated by experiment~\cite{Biercuk2009, BiercukPRA2009, Uys2009}, in order to describe dephasing in a generic qubit system.   We address only phase randomization due to classical environmental noise, described by a Hamiltonian written as
\begin{equation}
H=\frac{1}{2}[\Omega+\beta(t)]\hat{\sigma}_{Z},
\end{equation}
where $\Omega$ is the unperturbed qubit splitting, $\beta$ is a time-dependent classical random variable, and $\hat{\sigma}_{Z}$ is a Pauli operator.  For our purposes a full quantum mechanical treatment is not required, although it is possible following previous work~\cite{Uhrig2007,Uhrig2008,Lee2008, Cywinski2008}.  Further, as spin lifetimes in GaAs can exceed 1 s, current experiments are dominated by the effects of dephasing. 
\\
\indent The term $\beta(t)$ captures environmental fluctuations that produce an effective, fluctuating magnetic field on the qubit system~\cite{Kuopanportti2008,Cywinski2008};  all qubit-specific features of the model are captured through $\beta(t)$.  This term may appear due to external fluctuating magnetic fields, or intrinsic processes such as nuclear spin flips.
\\
\indent In the frame rotating at $\Omega$, $\beta(t)$ produces a random phase shift between the qubit basis states that on average leads to a $\sim1/e$ decay in coherence when the root-mean-squared phase accumulation is $\sim\sqrt{2}\pi$.  The characteristic timescale for this process is known as $\tau_{\phi}$, and in the absence of relaxation is equivalent to $T_{2}$, generically known as the decoherence time.  For the remainder of this manuscript we refer to the $1/e$ coherence time of the qubit as $T_{2}$.  
\\
\indent From an experimentalist's perspective it is useful to characterize $\beta(t)$ in the frequency domain, using the noise power spectrum $S_{\beta}(\omega)$, the Fourier transform of the two-time correlation function of $\beta(t)$,  
\begin{equation}
S_{\beta}(\omega)=\int_{-\infty}^{\infty}e^{-i\omega\tau}\left\langle\beta(t+\tau)\beta(t)\right\rangle d\tau.
\end{equation}
\noindent Here $\omega$ is angular frequency.  The influence of the noise term after time $\tau$ enters the measure of qubit coherence for a superposition state in the equatorial plane of the Bloch sphere by writing
\begin{equation} 
W(\tau)=|\overline{\langle\sigma_{Y}\rangle(\tau)}|=e^{-\chi(\tau)},
\end{equation}
\noindent where angled brackets indicate a quantum-mechanical expectation value, the overline indicates an ensemble average, and
\begin{equation}\label{eq:FF}
\chi(\tau)=\frac{2}{\pi}\int\limits_{0}^{\infty}\frac{S_{\beta}(\omega)}{\omega^{2}}F(\omega \tau) d \omega.
\end{equation}
In the expression above, $F(\omega\tau)$ is known as a filter-function \cite{Uhrig2007, Cywinski2008, Suter_Filter, Biercuk_Filter}  which encapsulates the experimental conditions under which qubit coherence is measured.  Since the filter function enters the coherence integral as a multiplicative factor of $S_{\beta}(\omega)$, small values of $F(\omega \tau)$ where $S_{\beta}(\omega)$ is large will lead to small values of $\chi(\tau)$, and hence coherence $W(\tau)\approx1$.

\subsection{Free-Induction Decay}
\indent In a Free-Induction-Decay (FID) experiment the filter function takes the form
\begin{equation}
F(\omega\tau)= 4 \sin^{2}(\omega\tau/2),
\end{equation}
such that qubit coherence is set by the integral over all spectral components of $S_{\beta}(\omega)$, when the small angle approximation is valid.  This intuitively makes sense, as in a FID experiment the precise spectral characteristics of the noise are immaterial - only the net average phase accumulation matters, consistent with the fact that the phase accumulation in the time domain is given by $e^{i\int_{0}^{t}\beta(t')dt'}$.  
\\
\indent Low-frequency fluctuations enter into an ensemble average measure of decoherence as they produce shot-to-shot quasistatic phase offsets at the conclusion of an experiment.  When averaged over many experiments these phase offsets produce a decay in coherence.  This form of decoherence is sometimes referred to as a measurement of $T_{2}^{*}$ in a time-ensemble-averaged fashion.

\subsection{Dynamical Decoupling Sequences}
\indent Dynamical decoupling is a technique derived from the Nuclear Magnetic Resonance (NMR) community that has been proven useful for suppressing decoherence in a quantum informatic setting \cite{Viola1998,Viola1999, Zanardi1999,Vitali1999, Byrd2003,Khodjasteh2005, Yao2007, Kuopanportti2008, Gordon2008, Biercuk2009, BiercukPRA2009}.   In a dynamical decoupling framework, the free evolution of a qubit is broken into discrete time periods between which time-reversing operations are applied, effectively decoupling the qubit from its fluctuating environment.  This approach is based on Hahn's discovery~\cite{Hahn50} of the spin echo in NMR for the mitigation of inhomogeneous dephasing, but applies equally well to the case of a single-spin.  
\\
\indent The quantitative effect of dynamical decoupling is captured by noting that the action of intermittent application of $\sigma_{Y}$ operators (``$\pi$''-pulses about the $Y$-axis) leads the phase accumulation, $e^{i\int_{0}^{t}\beta(t')dt'}$, to be broken into segments corresponding to the interpulse periods, with the sign of the phase accumulation alternating in successive periods.  Uhrig \cite{Uhrig2008} and Cywinski et al. \cite{Cywinski2008} showed that for any $n$-pulse sequence one may account for this by writing a time-domain filter function, $y_{n}(t)$, with values $\pm1$, alternating between each interpulse free-precession period.  We then write the filter function in the  frequency domain, $F(\omega\tau)=|\tilde{y}_{n}(\omega\tau)|^{2}$, where $\tilde{y}_{n}(\omega\tau)$ is the Fourier transform of the time-domain filter function, and $\tau$ is the total sequence length.  Again, the filter function describes phase accumulation in the frequency domain under the application of a dynamical decoupling pulse sequence.
\\
\indent For an arbitrary $n$-pulse sequence we may thus write
\begin{eqnarray}
&F(\omega\tau)=|\tilde{y}_n(\omega\tau)|^{2}\nonumber\\
&=|1+(-1)^{n+1}e^{i\omega\tau}+2\sum\limits_{j=1}^n(-1)^je^{i\delta_j\omega\tau}|^{2} 
\end{eqnarray}
where $\delta_{j}\tau$ is the time of the $j^{\rm th}$ $\pi_{X}$ pulse.  For convenience we assume instantaneous pulses, but we address this point in a later section of this manuscript (Section~\ref{sec:TauPi}).  
\\
\indent Using this model we may therefore determine the average dephasing error expected for any qubit given a well defined $S_{\beta}(\omega)$ and the analytically defined filter function for an arbitrary applied dynamical decoupling sequence.  This is an extremely efficient and practical model allowing all details of system-bath interactions (including external as well well as intrinsic microscopic noise processes) to be represented through the form of  $S_{\beta}(\omega)$.  This construction has been quantitatively validated by experiments using trapped atomic ions~\cite{Biercuk2009, BiercukPRA2009, Uys2009}, in which good agreement was found between experiment and theory for a wide range of $S_{\beta}(\omega)$ and applied pulse sequences.

\subsection{\label{subsec:CPMGvsUDD}Dynamical Decoupling Sequences of Interest}
\indent In this study we focus on the potential error-suppressing capabilities of two well studied sequences that have attracted significant attention in the last several years, CPMG and UDD.  These sequences form an instructive set to study due to their relative ease of implementation and distinct spectral-filtering characteristics~\cite{Uhrig2007, Uhrig2008, Suter_Filter, Biercuk_Filter}.
\\
\indent The $n$-pulse CPMG sequence is an extension of the spin-echo sequence and has been widely utilized in NMR and quantum information experiments~\cite{Haeberlen1976, Vandersypen2004}.  The sequence has been shown effective at suppressing phase randomization when noise processes are dominated by low-frequency components (e.g. $S_{\beta}(\omega)\propto1/\omega$)~\cite{BiercukPRA2009}.  An $n$-pulse CPMG sequence has all pulses evenly spaced, with the first and last free precession periods half as long as the interpulse free precession periods.  For the purposes of this study we do not differentiate between the Carr-Purcell and CPMG variants of the multipulse spin echo (CPMG is more effective at suppressing pulse rotation errors and most efficiently suppresses dephasing only when the initial Bloch vector is aligned with the applied transverse field).
\\
\indent Uhrig analytically derived~\cite{Uhrig2007} an $n$-pulse sequence in which the first $n$ derivatives of $\tilde{y}_{n}(\omega\tau)$ vanish for $\omega\tau=0$, exclusively through manipulation of the relative pulse locations, $\delta_{j}$.  The sequence is constructed using $\pi$ pulse locations determined analytically as $\delta_{j}^{(n)}= \sin^{2}[\pi j/(2n+2)]$, for an $n$-pulse sequence. The resulting sequence, UDD, has been shown to be extremely effective at suppression of high-frequency noise with a sharp spectral cutoff, appropriate for noise that may be present in, e.g. a spin-boson model.  The performance of UDD has been studied extensively theoretically and experimentally~\cite{Uhrig2007,Uhrig2008,Lee2008, Yang2008, Cywinski2008, Biercuk2009, BiercukPRA2009, Liu_Nature_UDD, Warren_UDD}.  
\\
\indent Both the CPMG and UDD sequences are capable of extending the qubit coherence time, but UDD has proven especially efficient at suppressing dephasing-induced error at times short compared to the qubit's $1/e$ decay time.  This regime is especially important for quantum information processing where constraints imparted by quantum error correction necessitate exceedingly low single-qubit error rates.
\\
\indent Additionally, the relative differences in the form of the filter functions of CPMG and UDD, and the commensurate performance differences under $S_{\beta}(\omega)$ dominated by either low-frequency contributions or high-frequency spectral components, make the use of these two sequences an interesting diagnostic tool for noise spectroscopy.  Quantitatively characterizing the expected dephasing error in a qubit under application of UDD and CPMG for different $n$ and $\tau$ will therefore assist in reconstructing the spectral characteristics of relevant noise sources.

\section{\label{sec:1f2}The Influence of $1/\omega^{2}$ Noise on Singlet-Triplet Qubit Coherence}
\indent A number of decoherence sources have been identified in S/T qubits, with dominant sources relating to the temporal evolution of nuclear spins in the GaAs heterostructure~\cite{Johnson_Nature2005, Taylor_PRB2007, ReillyOverhauser}.  A process known as spectral diffusion permits a zero-energy exchange process between distant nuclear spins, changing the Overhauser field experienced by the electrons, and thus causing dephasing.   Statistical measurements of the S/T qubit's singlet-return-probability suggest the presence of an effective noise power spectral density $S_{\beta}(\omega)\propto 1/\omega^{2}$.  While these nuclei can be treated as a fully quantum mechanical bath, it is sufficient to treat the fluctuating Overhauser field associated with the nuclei as a classical noise source producing dephasing.
\\
\indent We therefore begin with a study of the influence of $S_{\beta}(\omega)\propto 1/\omega^{2}$ on dynamical decoupling performance, informed by experimental observations.  We bound our spectral range of interest $\omega/2\pi \in[0.01, 10^{8}]$ Hz, in order to prevent numerical divergences.  The selection of these bounds is consistent with both experimental observations and general physical arguments.  Typical experiments on S/T qubits show $T_{2}^{(FID)}\approx$ 20-50 ns, and maximum coherence times with multipulse dynamical decoupling sequences $T_{2}^{(CPMG_{16})}\approx200\mu$s.  A noise contribution oscillating with frequency 0.01 Hz corresponds to a 100 s period, indicating that this contribution is static to roughly $10^{-6}$ over even the longest experiments ($\sim10^{-10}$ for a FID experiment).  We have confirmed in the forthcoming simulations that reducing this low-frequency bound further negligibly impacts the results.  
\\
\indent Similarly the high-frequency cutoff is motivated by bounds of the so-called dynamical decoupling limit.  If fluctuations in the noise are sufficiently rapid compared to the smallest interpulse spacing in an $n$-pulse sequence, $\omega_{max}/2\pi\gg\tau_{min}^{-1}$, then the application of a multipulse sequence can only lead to an \emph{increase} in net dephasing error.  Experiments on S/T qubits~\cite{Bluhm_LongCoherence, Barthel_Interlaced} have used up to $n\approx 20$, indicating a value of $\tau_{min}\approx 5\mu$s $=(200\;\textrm{kHz})^{-1}$ for CPMG.  The selected value of $\omega_{max}/2\pi\gg\tau_{min}^{-1}$, but the form of $S_{\beta}(\omega)$ ensures that components at these frequencies do not contribute significantly.

\begin{figure}[t]
\begin{center}
  \includegraphics[width=\columnwidth]{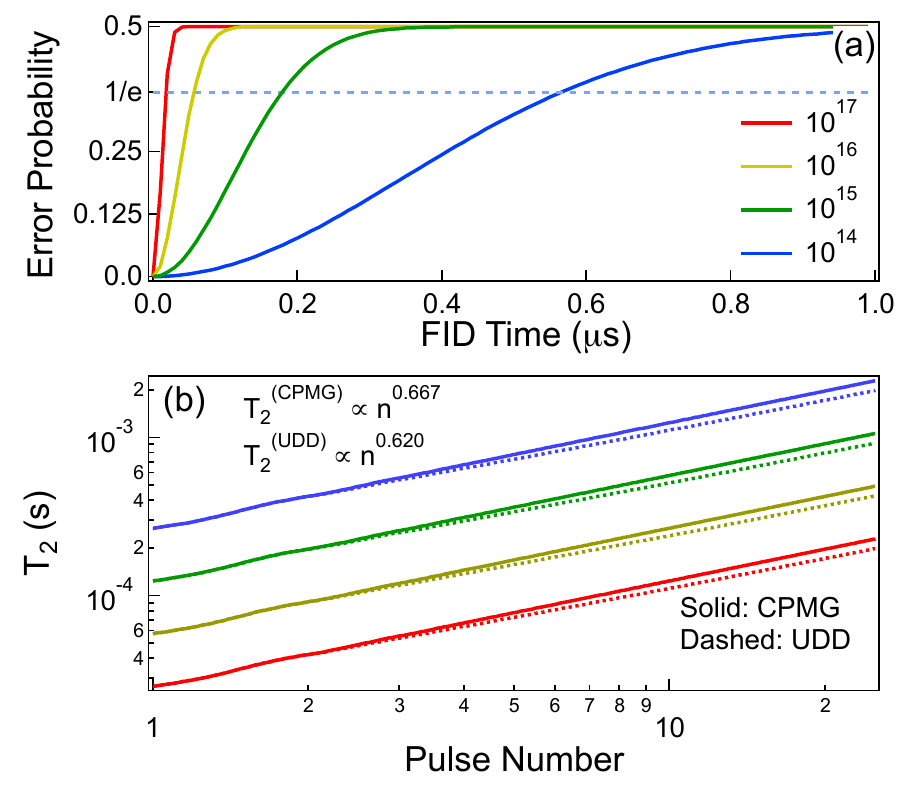}\\
\caption{\label{Fig:alpha}Qubit coherence in the presence of $1/\omega^{2}$ noise.  (a) Effect of noise scaling, $\alpha$, on predicted $T_{2}^{(FID)}$.  A value of $\alpha\approx5\times10^{16}$ most closely reproduces the observed $T_{2}^{(FID)}$.  The dashed line represents the $1/e$ error rate.  (b) Expected $1/e$ coherence time for CPMG and UDD pulse sequences as a function of pulse number for different values of $\alpha$. 
}
\label{default}
\end{center}
\end{figure}

\subsection{Calibrating the Noise Strength}
\indent We assess the scaling factor, $\alpha$, that serves to set the overall magnitude of the applied noise, writing $S_{\beta}(\omega)=\alpha \omega^{-2}$, such that for the minimum frequency defined within the spectral range above, $S_{\beta}(2\pi\;0.01\;\textrm{Hz})=\alpha$.  We determine the appropriate scaling by studying the calculated FID signal for various values of $\alpha$.  Physical insight into the approximate values of $\alpha$ is derived by considering the phase evolution in a FID experiment.  In the small angle approximation the ensemble averaged dephasing is given by $W(t)=exp[-\alpha(2/\pi)t^{2}\int_{0}^{\infty}S_{\beta}(\omega)d\omega]=exp\left[-\left(t/T_{2}^{(FID)}\right)^{2}\right]$.  This corresponds to a Gaussian decay where all factors (other than $t^{2}$) may be combined to give the $1/e$ decay constant, $T_{2}^{(FID)}$.
\\
\indent The $1/e$ coherence time of the FID signal is reduced exponentially with $\alpha$, as demonstrated in  Fig.~\ref{Fig:alpha}a.  Here we show the accumulated error (dephasing) signal as a function of $t$, the FID experiment time, for various values of $\alpha$.  Our measure of coherence is presented as $(1-W(t))/2$ a probability of qubit measurement in a selected basis state at the conclusion of the experiment.  A value of 0.5 corresponds to 50\% probability of the qubit being measured in  $\left|S\right\rangle$ or $\left|T\right\rangle$ --- complete dephasing.  
\\
\indent Comparison of simulations to experimental measurements gives an estimate of the noise scaling factor, $\alpha$.  First, we look to measured values of $T_{2}^{(FID)}$, and find values of $\alpha=10^{16}-10^{17}$ reproduce $T_{2}^{(FID)}\approx20-50$ ns with an approximately Gaussian decay.  This range of $\alpha$ gives a predicted value of $T_{2}^{(Echo)}\approx20-60\;\mu$s, corresponding to an improvement in coherence time of \emph{three orders of magnitude} using only a single pulse, and consistent with experiment~\cite{Bluhm_LongCoherence}
\\
\indent Further information about the appropriate selection of $\alpha$ can be derived by examining the calculated values of the $1/e$ coherence time for $n$-pulse dynamical decoupling sequences ($T_{2}^{(CPMG_{n})}$ and $T_{2}^{(UDD_{n})}$) (Fig.~\ref{Fig:alpha}b).  Comparison against experimentally measured values of $T_{2}^{(FID)}$, $T_{2}^{(Echo)}$, $T_{2}^{(CPMG_{6})}$, $T_{2}^{(CPMG_{10})}$, and $T_{2}^{(CPMG_{16})}$, narrow our approximation to $\alpha\approx5\times10^{16}$.  We note explicitly that this assumption is limited by measurement imprecision and variability between experiments, as well as limitations in our model.  
\\
\indent The value of $\alpha$ extracted from our simple phenomenological model can be compared with experimental measurements of the fluctuating nuclear field believed to be responsible for the $1/\omega^{2}$ form of the noise power spectral density.  Direct measurements of the noise spectrum around 1 Hz, where it already shows $1/\omega^{2}$ behavior independently leads to $\alpha\approx2.2\times10^{16}$, within a factor of order unity of the value extracted from our simulations.

\subsection{Coherence with Multipulse DD}
\indent The data presented in Figure~\ref{Fig:alpha}b provide information on the scaling of coherence time with $n$, and may be compared against full analytical calculations for spin-dephasing  and experimental measurements.  Qualitatively, we see that the spin-echo, CPMG, and UDD sequences all efficiently eliminate low-frequency contributions to dephasing.  Thus it is observed that application of even a single spin-echo pulse will significantly extend qubit coherence~\cite{Bluhm_LongCoherence}, and the application of sequences with increasing values of $n$ will provide diminishing returns.  Data for fixed $\alpha$ are well approximated by a power law scaling $T_{2}^{(CPMG_{n})}\propto n^{\psi^{(CPMG)}}$ or  $T_{2}^{(UDD_{n})}\propto n^{\psi^{(UDD)}}$ (indicated by an approximate linear scaling on a log-log plot).  Best fit values of $\psi$ for the two sequences are $\psi^{(CPMG)}=0.667$ and $\psi^{(UDD)}=0.620$.  These results are commensurate with analytical predictions of Ref.~\onlinecite{Cywinski2008} in which $\psi=\frac{2}{3}$ was predicted  in the presence of Gaussian noise.
\\
\indent In these calculations UDD is shown to generally yield shorter values of the measured $T_{2}$ relative to CPMG for this $S_{\beta}(\omega)$.  This is consistent with experimental measurements in a variety of systems, and is expected from a simple examination of the filter functions for CPMG and UDD; the UDD sequence trades an extremely small value of $F(\omega\tau)$ at low $\omega$ for a slightly lower ``turn-on'' frequency, above which the filter function does not suppress dephasing~\cite{Biercuk_Filter}.

\begin{figure}[tp]
\begin{center}
  \includegraphics[width=\columnwidth]{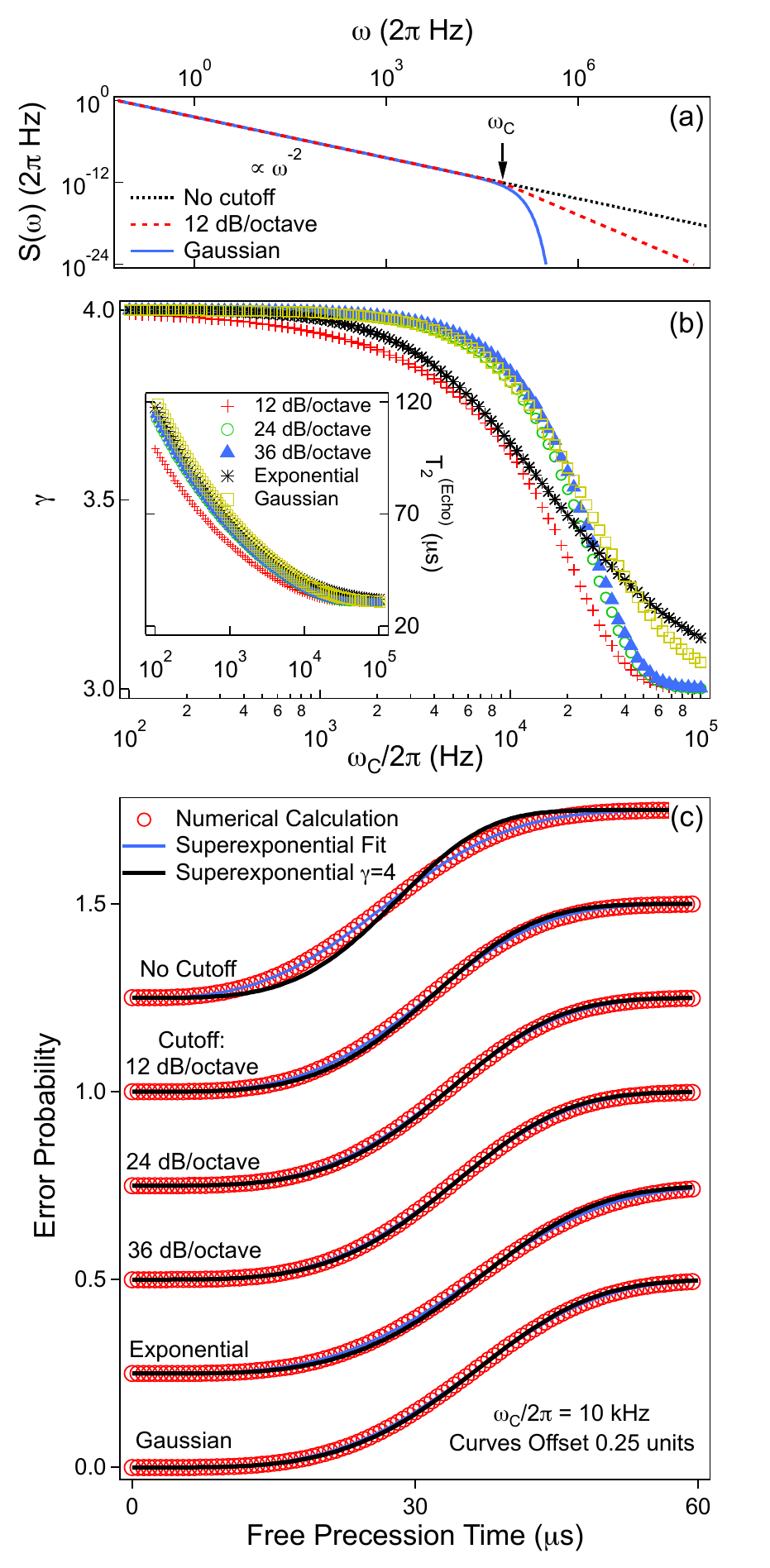}\\
\caption{\label{Fig:1f2cut1d} The effects of a high-frequency cutoff on calculated spin echo signal.  (a) Schematic of $S (\omega)$ including an high-frequency cutoff to account for the dynamics of nuclear flip-flops.    The high-frequency cutoff, $\omega_{C}$ is variable, as is the form of the rolloff above $\omega_{C}$. (b) $\gamma$ as a function of $\omega_{C}$ for different forms of the high-frequency rolloff.  (Inset) $T_{2}^{(Echo)}$ as a function of $\omega_{C}$ for the same parameters.  (c) Error probability as a function of free-precession time in a spin echo experiment.  Data approximated with a best fit superexponential: No cutoff, $\gamma$=3;  12 dB/octave, $\gamma=3.62$; 24 dB/octave, $\gamma=3.81$;  36 dB/octave, $\gamma=3.84$; Exponential, $\gamma=3.65$; Gaussian $\gamma=3.81$.  For these rolloffs $T_{2}^{(Echo)}\approx30-37\;\mu$s.  $\chi^{2}$ values for fits using $\gamma=4$: 12 dB/octave, $\chi^{2}$=0.010; 24 dB/octave, $\chi^{2}$=0.002;  36 dB/octave, $\chi^{2}$=0.002; Exponential, $\chi^{2}$=0.009; Gaussian $\chi^{2}$=0.001.  $\alpha=5\times10^{16}$. 
}
\end{center}
\end{figure}

\section{\label{sec:White}The High-Frequency Cutoff}
\indent The results presented thus far, and the similarity between these numerical calculations and data obtained by both experiments and more careful analytical approaches lend weight to the utility of this phenomenological approach.  In this section we will examine the potential impact of more carefully modeling the high-frequency nuclear spin dynamics.  The error accumulation in a FID experiment is again set by the integrated noise power spectral density, meaning that modification of the high-frequency contributions to $S_{\beta}(\omega)$ yields negligible changes.  Unlike the discussion of the low-frequency cutoff above, it is anticipated that the specific details of high-frequency noise will significantly impact the performance of the spin echo and multipulse dynamical decoupling
\\
\indent We consider modification of $S_{\beta}(\omega)$ to account for the high-frequency dynamics of the nuclear spin bath.  Above a correlation time $\tau_{C}=2\pi/\omega_{C}$, it has been calculated that spectral diffusion processes are suppressed.  This corresponds to a high-frequency cutoff, $\omega_{C}$ above which $S_{\beta}(\omega)$ is reduced below $1/\omega^{2}$ scaling.   We study the effects of various values of $\omega_{C}$ relative to experimentally relevant timescales (e.g. $T_{2}^{(Echo)}$, as well as different forms of the dynamics above $\omega_{C}$, and examine in detail the effects on qubit coherence.

\subsection{Spin Echo Decay}
\indent Beginning with the spin echo we study the impact of $\omega_{C}$ and the form of the high-frequency rolloff on the measured qubit coherence. Experimental data for coherence as a function of free-precession time have been shown to be fit well to a superexponential decay $\propto exp(-(t/T_{2})^{\gamma})$, with $\gamma=4$.  We numerically calculate the spin-echo decay as a function of free-precession time and fit to a superexponential for various values of $\omega_{C}$, and different high-frequency rolloffs.  As seen in Fig.~\ref{Fig:1f2cut1d}b, varying $\omega_{C}$ has a significant effect on the best-fit values of $\gamma$ extracted from the numerical calculations.  As $\omega_{C}$ increases we observe a reduction of $\gamma=4\rightarrow3$, corresponding to the presence of increased high-frequency spectral weight in $S_{\beta}(\omega)$. For the smallest values of $\omega_{C}$ the $1/e$ coherence time expected for a spin echo experiment increases dramatically beyond what has been observed.  Offsetting these changes to the predicted $T_{2}^{(Echo)}$ by increasing $\alpha$ would be inconsistent with microscopic modeling and would yield values of $T_{2}^{(FID)}$ inconsistent with experiment, thus limiting the range of possible values of $\omega_{C}$.  These calculations suggest a possible value of $\omega_{C}/2\pi\approx$ 10-100 kHz to best approximate experimental measurements.  Such values are in line with analytical calculations of the high-frequency cutoff for nuclear-spin diffusion processes.
\\
\indent The form of the high-frequency cutoff in $S_{\beta}(\omega)$ also has significant impact on the experimentally measured qubit coherence.  A microscopic, quantum mechanical treatment predicts that CDD and UDD sequences can decouple the electron-nuclear interaction to arbitrary order \cite{Witzel2007, Lee2008}. These results already indicate that (to the extent that a description in terms of a noise spectral density is valid) the spectral density should have a high frequency cutoff that is faster than any power law. A $\omega^{-\zeta}$ cutoff would lead to a $\exp(-t^{\zeta+1})$ decay for UDD or CDD decoupling sequences of sufficiently high order according to a perturbative short-time expansion.  We argue that a Gaussian rolloff where the noise power spectrum is suppressed as  $S_{\beta}(\omega)\propto \omega^{-2} e^{-(\omega/\omega_{C})^{2}}$ is most appropriate for the dynamics of nuclear spin diffusion, motivated by the the pair correlation approximation of Ref.  ~\onlinecite{Yao2006} (see Appendix for details).  
\\
\indent We present numerical simulations assuming this form of high-frequency rolloff, as well as power law scaling $\omega^{-\zeta}$, corresponding to a rolloff of $3\zeta$ dB/octave (e.g. 12 dB/octave = $\omega^{-4}$). 
The transition from $\gamma=4\rightarrow3$ as a function of $\omega_{C}$ varies with the form of the high-frequency rolloff (Fig.~\ref{Fig:1f2cut1d}b).  Setting $\omega_{C}/2\pi=10$ kHz in Fig.~\ref{Fig:1f2cut1d}c, we show full numerical calculations for accumulated error in a spin-echo experiment as a function of free-precession time, superexponential fits to these data, and superexponential fits with $\gamma$ fixed to be four, for different values of the high-frequency rolloff. We observe that for the steepest values of the rolloff the calculated error rates are fit well using $\gamma\approx4$, while reducing the steepness of the rolloff (corresponding to increasing the weight of high-frequency components in $S_{\beta}(\omega)$) suppresses the best-fit $\gamma$ towards three.  Calculations using a Gaussian rolloff show good agreement with a superexponential decay using $\gamma=4$.  As shown in panel b, however, these differences are reduced further as $\omega_{C}\rightarrow100$ kHz.

\subsection{Multipulse Dynamical Decoupling}
\begin{figure*}[htp]
  \includegraphics[width=12cm]{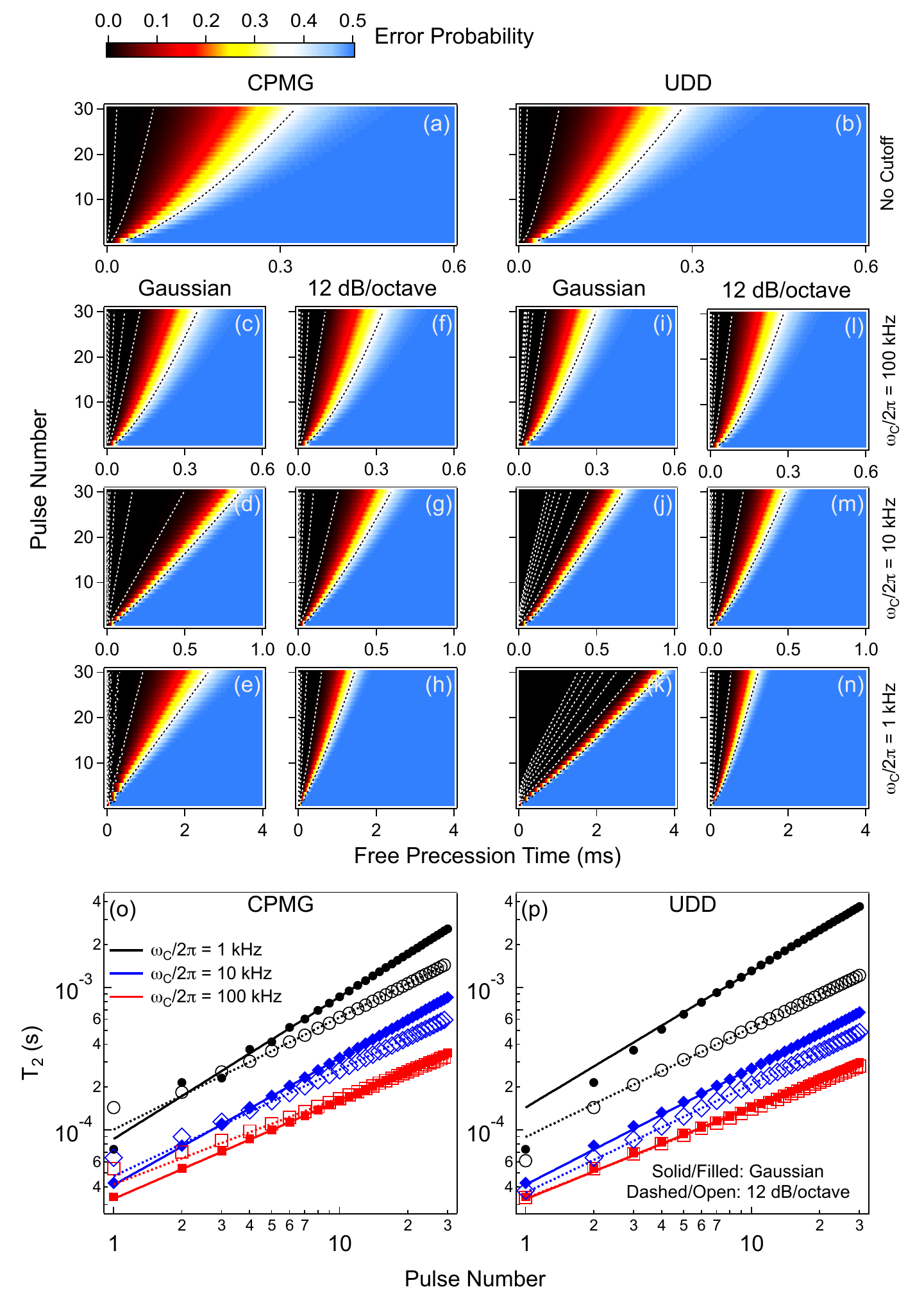}\\
\caption{\label{Fig:1f2cutoff}The effect of a high-frequency cutoff in the $1/\omega^{2}$ noise power spectrum on qubit coherence under the application of multipulse dynamical decoupling.  Qubit error probability (coherence) is expressed as a colorscale, with a value of 0.5 corresponding to complete dephasing.  Left panels correspond to CPMG, right panels correspond to UDD.  Individual columns for panels (c-n) denoted with high-frequency rolloff in use. Black dashed line corresponds to the $1/e$ decay contour.  White dashed lines correspond to constant error contours beginning at 1\% and diminishing by 100$\times$ per step, from right to left.  Contours thus correspond to error probability $10^{-2}$, $10^{-4}$, $10^{-6}$, etc.  Contours for values of error probability below $10^{-12}$ are not shown. (o-p) Extracted $T_{2}$ for CPMG (o) and UDD (p) given different values of $\omega_{C}$ and the high-frequency rolloff.  Lines are power-law fits to the data holding the $n=0$ intercept fixed at $T_{2}^{(FID)}=26$ ns, and varying $\psi$.  Extracted fit parameters: Gaussian Rolloff - $\psi_{\textrm{1 kHz}}^{(UDD)}=0.956$, $\psi_{\textrm{10 kHz}}^{(UDD)}=0.817$, $\psi_{\textrm{100 kHz}}^{(UDD)}=0.647$, $\psi_{\textrm{1 kHz}}^{(CPMG)}=0.999$, $\psi_{\textrm{10 kHz}}^{(CPMG)}=0.891$, $\psi_{\textrm{100 kHz}}^{(CPMG)}=0.692$; 12 dB/octave Rolloff -  $\psi_{\textrm{1 kHz}}^{(UDD)}=0.769$, $\psi_{\textrm{10 kHz}}^{(UDD)}=0.762$, $\psi_{\textrm{100 kHz}}^{(UDD)}=0.628$, $\psi_{\textrm{1 kHz}}^{(CPMG)}=0.788$, $\psi_{\textrm{10 kHz}}^{(CPMG)}=0.748$, $\psi_{\textrm{100 kHz}}^{(CPMG)}=0.608$;.  For these calculations, $\alpha=5\times10^{16}$.}

\end{figure*}

\indent Divergences between the superexponential fit with $\gamma=4$ and full numerical calculations show differences that may be observable in experiments, but differences in the form of the spin-echo decay are subtle and will likely be limited by experimental noise.  The scaling of qubit coherence times and error rates in the high-fidelity regime with pulse number, $n$, provide further insight into the high-frequency nuclear dynamics.  
\\
\indent In Fig.~\ref{Fig:1f2cutoff} we show the calculated error rates in colorscale as a function of free-precession time and $n$, for CPMG and UDD.  We vary the value of $\omega_{C}$ and the form of the high-frequency rolloff between the ``hard'' Gaussian and the ``soft'' 12 dB/octave power-law to serve as performance bounds.  In the high-fidelity regime constant-error contours demonstrate the efficiency of error suppression for times short compared to $T_{2}$.   As $\omega_{C}$ is reduced calculated coherence times increases in all circumstances.  In the case of a hard Gaussian rolloff, the form of the UDD filter function~\cite{Uhrig2008} provides significant benefits relative to CPMG and provides much deeper error suppression as well as enhanced coherence times.  By contrast, the presence of a slow 12 dB/octave rolloff suppresses any benefits of using UDD relative to CPMG.
 \\
 \indent Aside from the absolute magnitudes of the extracted coherence times, the high-frequency components of $S_{\beta}(\omega)$ also impact the form of the scaling of the calculated error suppression with $n$.  For small values of $\omega_{C}$ the addition of pulses produces an approximately linear increase in coherence time with $n$, as high-frequency fluctuations contribute little dephasing.  For $\omega_{C}\approx1-10$ kHz, coherence times in the ms regime are possible for $n\approx20$.  For larger values of $\omega_{C}$ coherence times of this order are possible with increasing $n$, but as discussed previously, the form of $S_{\beta}(\omega)$ yields diminishing returns with $n$.  The highest frequency components of interest are set approximately by $\tau_{min}^{-1}$, the minimum interpulse free-precession time.  When  $\tau_{min}^{-1}\gg\omega_{C}/2\pi$, reduction in $\tau_{min}$ (via increasing $n$) fails to provide additional measurable suppression of noise.  However, in the case of relatively large values of $\omega_{C}$ small improvements in measured error rates may be obtained with increasing $n$. 
 \\
 \indent These characteristics are captured in the power-law scaling of $T_{2}^{(CPMG_{n})}\propto n^{\psi^{(CPMG)}}$ ($T_{2}^{(UDD_{n})}\propto n^{\psi^{(UDD)}}$) Fig.~\ref{Fig:1f2cutoff}o-p shows the effect of the high-frequency cutoff on $\psi$.  For both CPMG and UDD lowering the value of $\omega_{C}$ increases $\psi\rightarrow1$.  However, this phenomenon is most pronounced in the case of a hard Gaussian rolloff; the presence of a soft high-frequency rolloff suppresses the changes in $\psi$ with $\omega_{C}$.

\subsection{Low-Frequency Nuclear Dynamics and Programming}
\indent For the sake of a transparent and self-contained presentation, we have assumed that the spectrum is well described by a $1/\omega^{2}$ power law down to a suitable low-frequency cutoff.  A full treatment of nuclear spin diffusion shows that this approximation breaks down below $\sim$0.1 Hz;  below this frequency the power spectral density shows significant deviations from $1/\omega^{2}$ scaling.  However, these frequencies only affect the relation between $\alpha$ and $T_{2}^{(FID)}$, whereas the behavior at higher frequencies relevant for decoupling sequences remains unaffected.  The direct determination of $\alpha$ as described above provides an independent estimate that is not affected by the specific form of the low-frequency spectral density.  Further, calculations including modified low-frequency behavior show that the best-fit value of $\alpha$ remains in the range $10^{16}-10^{17}$.
\\
\indent From the calculations presented heretofore it is readily apparent that one of the most effective ways to improve qubit coherence is to mitigate the fluctuating overhauser field experienced by the qubit.  Nuclear spin diffusion has been characterized and even controlled~\cite{Reilly_Science_Nuke, Bluhm_Feedback} through feedback mechanisms, allowing the time-ensemble-averaged $T_{2}^{*}$ to be extended.  The most important parameter in such an experiment is the \emph{bandwidth} of the noise suppression procedure; mitigating only low-frequency noise will not improve coherence limits with $n$-pulse dynamical decoupling sequences where the limiting performance is governed by rapid fluctuations at frequencies near the dynamical decoupling limit.  For $S_{\beta}(\omega)\propto 1/\omega^{2}$ we find that using experimentally relevant parameters, suppression of the low-frequency tail has a dominant effect exclusively on $T_{2}^{(FID)}$ unless the bandwidth of the noise suppression extends beyond 100 kHz.  As such, programming nuclear dynamics to mitigate low-frequency fluctuations can make $T_{2}^{(FID)}\rightarrow T_{2}^{(Echo)}$.  By contrast, in the presence of a nuclear spectrum with a sharp cutoff, if the bandwidth of nuclear stabilization far exceeds the cutoff frequency, $S_{\beta}(\omega)$ is suppressed over its entire relevant range, as given by the product of $S_{\beta}(\omega)$ and the filter transfer function defined by the nuclear programming protocol.  In this case, high-bandwidth nuclear stabilization has the potential to dramatically extend coherence and improve the efficacy of dynamical decoupling sequence performance.

\section{\label{sec:TauPi}Noninstantaneous Control Pulses}
\indent Realistic experiments are conducted with control $\pi$ pulses of nonzero duration ($\tau_{\pi}\neq0$), mitigating the utility of the so-called bang-bang limit in which most dynamical decoupling studies are performed.  In previous work, a simple modification of the filter function was implemented in order to account, to lowest order, for the effect of noninstantaneous control pulses~\cite{BiercukPRA2009}.  In this procedure, we modify the time-domain filter function by incorporating a delay with length $\tau_{\pi}$ and value zero between free-precession periods.  This approximation assumes that dephasing is negligible during the application of a $\pi$ pulse, consistent with many experimental observations.  Incorporating this delay results in a filter function for an arbitrary $n$-pulse sequence
\begin{eqnarray}
&F(\omega\tau)=|\tilde{y}_n(\omega\tau)|^{2}\nonumber\\
&=|1+(-1)^{n+1}e^{i\omega\tau}+2\sum\limits_{j=1}^n(-1)^je^{i\delta_j\omega\tau}\cos{\left(\omega\tau_{\pi}/2\right)}|^{2} \nonumber\\
&\;
\end{eqnarray}
where $\delta_{j}\tau$ is the time of the \textit{center} of the $j^{\rm th}$ $\pi$ pulse, and $\tau$ is the sum of the total free-precession time and $\pi$-pulse times.  In previous experiments we showed that this approximation could account for evolution of the noise field during the control pulses, improving the accuracy of the theoretical model.
\\
\indent We have employed this model in the current setting to understand the effect of the finite \emph{bandwidth} of the applied pulses in experiments on the singlet/triplet qubit.  Modeling $\tau_{\pi}\in$ [0.1,20] ns, we find negligible effects on qubit coherence (changes $<10\%$ in the calculated $1/e$ coherence time) for all spectra of interest (not shown).  These changes are largest for high values of the high-frequency cutoff ($\omega_{C}\approx100$ kHz), and typically less than 1\% for $\omega_{C}\approx10-30$ kHz.  Further, the finite bandwidth of the pulses did not produce discernible changes in the relative performance between the CPMG and UDD sequences.  Typical deviations in the high-fidelity regime demonstrated that the minimum accumulated error could be impacted at the $10^{-6}$ level for $\tau_{\pi}\approx100$ ns and $\omega_{C}>50$kHz.  These differences are not detectable in current experiments, suggesting that at present the effects of pulse-bandwidth are negligible.
\\
\indent The model presented above assumes ideal, but noninstantaneous control pulses.  A significant source of error in dynamical decoupling experiments may be derived from imprecise control operations.  The presence of a large error even for small $n$ in experiments on S/T qubits suggests that pulse-fidelities may be quite poor.  One expects that the effects of control pulse imperfection can be elucidated by experimental studies of error scaling with $n$; short-time behavior would show large error increases with $n$ in the presence of substantial pulse errors that increase with each applied operation.  Examining the relative performance of CP vs CPMG multipulse spin echo could also serve to illuminate the role of imperfect control pulses.

\section{\label{sec:Discuss}Discussion and Conclusion}
\indent In this manuscript we have presented a phenomenological model for the error accumulation in FID, Spin Echo, and multipulse dynamical decoupling experiments on Singlet-Triplet spin qubits in GaAs.  Our simulations are based on an experimentally validated model for error accumulation in the presence of a noisy environment characterized by an arbitrary $S_{\beta}(\omega)$.  The results presented herein suggest that a simple semiclassical noise model can accurately reproduce a variety of experimental measurements, and can provide strong agreement with more detailed theoretical calculations.  Further, they have elucidated some of the performance limits one might expect based on relevant noise processes associated with nuclear spin dynamics.
\\
\indent The salient characteristics of experiments~\cite{Bluhm_LongCoherence} providing the best measured values of $T_{2}^{(Echo)}$ and $T_{2}^{(CPMG_{n})}$, are reproduced by simulations incorporating $S_{\beta}(\omega)\propto 1/\omega^{2}$ in our phenomenological model.  An extraction of of the relevant noise strength, $\alpha$ based on comparison with these measurements and $T_{2}^{(FID)}$ agrees within a factor of order unity with more detailed calculations of nuclear-spin diffusion.  Further, our calculations provide data supporting the presence of a high-frequency cutoff $\omega_{C}\approx 100$ kHz with Gaussian rolloff,  consistent with heuristic microscopic theory and experimental measurements.  
\\
\indent The results presented here provide predictive and analytical power for experimental studies of multipulse dynamical decoupling, permitting detailed characterization of the form of $S_{\beta}(\omega)$ via observation of the scaling of $T_{2}^{(CPMG_{n})}\propto n^{\psi^{(CPMG)}}$ ( $T_{2}^{(UDD_{n})}\propto n^{\psi^{(UDD)}}$).  Calculations incorporating a high-frequency cutoff $\omega_{C}\approx 100$ kHz are consistent with experimental measurements~\cite{Marcus2010} of $\psi\approx0.7$.  Detailed information on the form of the high-frequency rolloff can be obtained by comparison of the performance of CPMG and UDD in the high-fidelity regime.  Unfortunately, at this time measurement and operational infidelities mask the effects of decoherence-induced error accumulation to be studied in this regime.
\\
\indent Achieving the ultimate benefits associated with UDD or other optimized dynamical decoupling protocols will thus require significant improvements in operational fidelity, measurement fidelity, and classical noise filtration.  The dynamics of nuclear spins likely provide the dominant dephasing mechanism in Singlet-Triplet qubits realized today, but as experimental capabilities improve, other noise sources will come into play.  If, for instance, it is demonstrated clearly that nuclear spin dynamics exhibit a high-frequency cutoff with a Gaussian rolloff, for frequencies $\omega>\omega_{C}$, an additional \emph{noise floor} will likely come into play due to other processes (e.g. ambient magnetic field fluctuations due to current noise near the sample).  Simulations show that including such a noise floor is effectively the same as modifying $S_{\beta}(\omega)$ to include a soft rolloff above the value of $\omega_{C}$ set by nuclear spin dynamics. These high-frequency spectral components limit coherence and reduce the benefits associated with the use of optimized decoupling strategies such as UDD.  Understanding system performance with soft noise rolloff therefore serves as a simple approximation for the ultimate influence of residual extrinsic noise processes.  We have also shown that detailed, systematic studies of S/T qubit coherence under the application of various pulse sequences and values of $n$ can provide a useful diagnostic for relevant noise sources, as an indirect form of noise spectroscopy~\cite{Cywinski2008}.  
\\
\indent These simulations have demonstrated that the most effective way to improve qubit coherence is through suppression of high-frequency noise. The difference in calculated coherence times and error rates associated with soft and hard cutoff frequencies motivates effort in hardware engineering in order to suppress any effects of external noise sources.  For instance, in the presence of weak, high-frequency-dominated electrical noise, improved suppression arising from a transition between single-pole and multi-pole filters in an experimental system could lead to coherence-time extension by 2-3$\times$ in a dynamical decoupling sequence.  Limiting dephasing noise to intrinsic sources could more importantly suppress error rates by orders of magnitude in the high-fidelity regime via use of UDD relative to CPMG. 
\\
\indent Historically, a large focus of the community has been on the realization of spin qubits in materials dominated by zero-nuclear-spin isotopes such as carbon or silicon.  The studies presented here support this general viewpoint as we believe the dynamics of nuclear spins in the GaAs host material to be the dominant source of observed dephasing.  However, alternate noise sources such as fluctuations in the effective exchange coupling due to electrical voltage noise may in fact dominate such experiments, and will occur irrespective of the presence of a nuclear spin bath.  If access to high-fidelity single-qubit operations is readily available, dynamical decoupling pulse-sequence application may form an effective means to reduce performance gaps between different materials systems, extending coherence by \emph{orders of magnitude} relative to $T_{2}^{(FID)}$.

\appendix*
\section{Gaussian High-Frequency Rolloff}
\indent The dynamics of nuclear spins are driven by the Knight shift due to the hyperfine interaction with the electron spin and interactions with neighboring nuclei. The latter drive flip-flop (i.e. spin transfer) processes between nuclei of the same species, which over long time scales cause the spin diffusion discussed above.  Inter-species flip-flops are suppressed by the mismatch in Zeeman energies due to the applied field. We argue that transitions of a single nuclear spin are driven by the transverse component of the (effective) interaction fields generated by nearby nuclei of the same species. The distribution of this field should be similar to the longitudinal component reflected in the NMR resonance line, but a factor 2-3 smaller since only the same-species resonant contributions are relevant here.  Based on the measured NMR line shape and the fact that many randomly oriented spins contribute, we assume a Gaussian distribution.
\\
\indent The Knight shift due to the hyperfine interaction with the electron spin is on the order of 10 G. However, its effect on the dynamics is much smaller since the electronic wave function extends over many unit cells so that nearby nuclei experience approximately the same Knight shift.  Only the Knight shift difference is relevant because it detunes the flip-flop interaction.  Taking a typical length scale of 5 nm over which the wave function changes appreciably along the z-direction, the Knight shift variation between nuclei that are one lattice constant
apart is on the order of 1 G, comparable to the transverse field setting the off diagonal matrix elements. Neglecting it is thus reasonable for the purpose of developing a rough phenomenological model.
\\
\indent Furthermore, it should be noted that the change of the Overhauser field from each such a spin flip-flop is much smaller than that resulting from flipping a single nuclear spin, since the angular momentum is only transfered over short distances to other nuclei with approximately the same hyperfine coupling constant. Instead of using the above estimate of this reduction based on the wave function shape, we simply approximate the Gaussian distribution of flip-flop frequencies acts as a multiplicative factor on the diffusion spectrum discussed above.  This phenomenological approach automatically solves the problem that our argument is not valid at longer time scales, when the diffusive process associated with many subsequent flip-flops sets in. At the corresponding low frequencies, the Gaussian factor is near unity, so that the diffusion spectrum is recovered.

\begin{acknowledgments}
We thank S.D. Bartlett, A.C. Doherty, C.M. Marcus, and D.J. Reilly for useful discussions. Thanks also go to H. Uys and J.J. Bollinger for the development of software for modeling qubit coherence.  This research was partially supported by the U.S. Army Research Office and the School of Physics, University of Sydney.
\end{acknowledgments}

\end{document}